\def\aj{{AJ}}

\def\apj{{ApJ}}
\def\apjs{{ApJS}}

\def\mnras{{MNRAS}}
\def\nat{{Nature}}

\def\apjl{{ApJLett}}
\def\apss{{APSS}}
\def\Nature{{Nature}}

\documentclass[cup5b]{caps}
\usepackage{graphicx}
\usepackage{amssymb}
\usepackage{ociwsymp3}   

\HeadText{B. Moore}  

\begin{document}

\pagenumbering{arabic}

\author[]{Ben Moore\\Institute for Theoretical Physics, University of Z\"urich, Switzerland}

\chapter{Evolutionary processes in clusters}

\begin{abstract}

Are the morphologies of galaxies imprinted during an early and rapid formation epoch 
or are they due to environmental processes that subsequently transform galaxies between 
morphological classes? Recent numerical simulations demonstrate that
the cluster environment can change the morphology of galaxies, even at a couple 
of cluster virial radii. The 
gravitational and hydrodynamical mechanisms that could perform such 
transformations were proposed in the 1970's, before the key 
observational evidence for environmental dependencies - the 
morphology-density relation and the Butcher-Oemler effect.

\end{abstract}

\section{Introduction}

Galaxies are observed to have a wide range of 
morphologies and stellar configurations,  classified as
disk-like with subclasses depending on the degree of disk instability, gas fraction 
and central nucleation, or spheroidal configurations of varying shapes and concentrations.
Amongst both of these broad sequences we have various combinations of irregularities, 
subclasses, sizes, luminosities and star formation histories.
Observational studies of clusters and groups have played an important role
in helping us to understand the origin of galactic morphologies.
This is due to three main reasons: (i) evolutionary processes are accelerated in
high density environments, (ii) some classes of galaxies are only found within 
larger virialised systems and (iii) clusters
can be found easily at higher redshifts and therefore can be used to
detect evolution directly. 

The observational data are consistent with the idea that the visible baryons 
are concentrated at the center of much larger dark matter halos (Fischer etal 2000). 
Thus the interpretation of galaxy morphologies is closely linked to understanding
dark matter clustering on different scales.
The baryons are observed to have a scale length that is about
1/10th of that of the dark matter implying that dissipation must have played
a key role in galaxy formation.
On average, galactic mass halos have accumulated close to the universal baryon fraction 
implying that violent feedback that leads to mass ejection of baryons has played a
less important role in determining galaxy morphology.
If most of the baryons quietly dissipated with little merging 
between subhalos, then the first galaxies to form are expected to be disks due to
conservation of angular momentum as the gas radiates energy, sinks within the dark matter
halos and spins faster.
Indeed, most of the galaxies in quiet environments (outside of other virialised systems) 
are disks - nearly all other classes of galaxies are found orbiting inside deeper potentials
as satellite galaxies (halos within halos).

Spheroidal stellar configurations span a factor of $10^7$ in luminosity from
central cluster cD's to the local group dwarf spheroidals - whereas the disks
range from the giant LSB's to the tiny Local Group dIrr's.
In general the spheroidals reach higher masses and luminosities. It is not
obvious why this is the case - the gas cooling timescale limits the maximum size of
cold baryonic systems but it is a puzzle why we do not 
observe $\sim$30kpc disks at the centre of 
clusters where the cooling times are short. Perhaps harassment or 
local feedback from a central AGN may be suppressing 
central disk formation in massive halos.

The fact that few spiral galaxies are found anywhere 
within the central regions ($\sim$Mpc) of rich 
clusters (Dressler 1980) could be explained in two ways:
(i) cluster galaxies were never disks since they formed in a more merger prone environment
as lenticulars (S0), spheroidals (E) or dwarf ellipticals (dE), 
or (ii) disks formed first and have subsequently been transformed into
other morphological classes by virtue of the cluster environment that formed later.
The Hubble Space Telescope allowed a direct observational test of the morphological
change over the past few Gyrs within dense and proto-dense environments. High resolution images
of  the ``Butcher-Oemler'' (Butcher and Oemler 1978)
clusters at $z\sim0.5$ showed that  the luminous spheroidal population 
was already in place at this epoch but that the majority of the other galaxies 
were indeed disks (Dressler etal 1997), and even the S0 population is 
difficient (Smail etal 1997).

Many of these distant Butcher-Oemler clusters
are complex merging systems of groups
that will eventually have similar masses and galaxy densities to
the nearby rich clusters like Coma. Clusters at high redshift that already 
have the mass and virial state of a rich cluster may already look 
similar to Coma, for example MS1054 (van 
Dokkum etal 1998).
These comparisons are complex due to projection effects and background subtraction, but
also the interpretation is difficult since frequently
one is comparing systems identified at different epochs that
are in different states of virialisation.

\section{The paradigms for disk and spheroid formation}

In order for gas to concentrate and form stars at the centers of dark halos
it must first be shock heated so that it can dissipate and cool to high densities. As
it cools it must spin faster as it conserves is primordial angular momentum generated 
from tidal torques (Hoyle 1949).
The natural end state of cooling gas within an isolated DM halo is
a rotationally supported disk; thus one might postulate that disks 
are the initial building block from which the entire morphological sequence is constructed.
Once the first disky objects have formed (or even whilst they are forming), 
a process of multiple mergers and the
associated central gas inflows are expected to create the cD, E sequence 
that extends to the faintest ellipticals - the high surface brightness M32-like systems.

It should be remembered that this is primarily 
theoretical speculation - the details of this process
are far from being worked out and considerable numerical resources are required 
to fully investigate these ideas. 
Firstly  we need to understand how the gas is shock heated and cools to the 
central disk and how the angular momentum of the gas evolves during this process.
Forming a bulgeless late type disk galaxy may be one of the most difficult challenges 
for the CDM model. Understanding the formation of spheroids has equally
challenging problems.
For example, why is there such a narrow spread in the luminosity
of cD galaxies and why do they lie offset brightwards from the galaxy luminosity function?
Why are the smallest ellipticals rotating faster than the massive ellipticals? Why are there so
few isolated field ellipticals that are the probable 
end state of the $M_*$ groups, where $M_*$ is the
characteristic non-linear mass today? The kinematics, colours and ages of 
ellipticals are also challenges for this paradigm that have yet to be understood.
On a more general note, it is still unclear how CDM type models that predict a
steep mass spectrum of halos can produce a flat luminosity function, whilst at
the same time matching the correlation between baryonic mass/luminosity and 
dynamical mass.

Independent of the model, we can pose the question whether it is 
theoretically possible to reproduce the entire sequence of galaxies 
starting from disky systems?
Starting with Holmberg's (1948) N-body experiments with lights and
photometers and later confirmed by Toomre's computer calculations (Toomre 1977), 
it has been shown that it is possible for galaxies to interact gravitationally and produce
spectacular tidal features.  Longer numerical calculations performed by
Gerhard (1981) showed that the end states of mergers will
violently relax into spheroidal configurations, but
additional dissipation is required to produce the high phase space
densities observed in ellipticals. As is usual with N-body
simulations, the detailed comparisons can be quite complex and the
more details one simulates the more discrepancies one finds between
data and theory (Naab etal 1999).  More realistically, most ellipticals
probably formed from the rapid and multiple mergers of a variety of
baryonic systems - not too unlike a clumpy monolithic collapse and
distinguishing between the two standard hypotheses is difficult. However
in this review I will concentrate on ``non-merging'' mechanisms that
can drive evolution and transform galaxies between and across
morphological classes in the environments of galactic, group and
cluster halos.

\section{Mechanisms for transformation}

Early theoretical work predicted that clusters of galaxies are harsh environments for
galaxies to inhabit. Hydrodynamical processes (Spitzer \& Bade 1950, 
Gunn \& Gott 1972, Cowie \& Songalia 1977,  Norman \& Silk 1979, Nulsen 1980)
 were proposed as important mechanisms for 
stripping the interstellar medium from galaxies. 
The importance of gravitational encounters and tidal forces as a mechanism for
forming central cD galaxies, creating diffuse light and for influencing morphological
transformation was proposed by many authors in the 1980's
(Gallagher \& Ostriker 1972, Ostriker \& Tremaine 1975, Richstone 1976,
White 1976, Hausman \& Ostriker 1977, Merritt 1983). 

Many of these mechanisms are efficient only in massive galaxy clusters and 
they may be invoked to explain the morphology density
and Butcher-Oemler effects (Solanes etal 1992, Kauffman 1995). 
However the role of evolution in lower density environments may play an
important role. The evolution 
of bright/massive galaxies within 
groups and poor clusters with dispersions below 400 km/s should be dominated by 
dynamical friction and mergers. An important question remains - is galaxy evolution
in clusters driven by pre-processing of galaxies in groups?  
Indeed, star formation appears to be truncated within galaxies that lie in
lower density environments (Balogh etal 2000).
However, it is clear that  galaxies in virialised systems have different morphologies than the field. 
If environment is responsible for transition, then it should be possible 
to quantify observationally and theoretically where, when and how 
these transitions take place.

In his excellent book on clusters, Sarazin (1980) makes six points to 
argue that the morphologies of galaxies are
set at the time of formation rather than by subsequent environmental processes.
Many of these points are still open questions. For example, Dressler (1980) claimed that
the bulges of S0's are more luminous than those of the spirals -- a frequently cited 
result that was disputed
by Solanes etal (1989) who analysed the entire Dressler sample and took into account additional 
selection effects. Most of Sarazin's 
arguments were made with the idea that gas processes would be driving a morphological
evolution  -- Sarazin makes the point that gas dynamics would not reproduce 
the thick disks of S0's, however it is a natural outcome of gravitational interactions.

Indeed, recent numerical work has demonstrated that gravitational and hydrodynamical 
processes could be responsible for many of the observed aspects of galaxy morphology
and evolution in different environments 
(Byrd \& Valtonen 1990, Valluri \& Jog 1991, Valluri 1993, Summers etal 1995, 
Moore etal 1996, Dubinski 1998, Moore etal 1998, Abadi etal 1999, Dubinski etal 1999, 
Mihos 1999, Quilis etal 2000, Vollmer etal 2000, Balogh etal 2000, Mayer etal 2001, 
Vollmer etal 2002, Gnedin 2003a, 2003b). The following sections will discuss these studies in the 
context of understanding the wide variety of galactic morphologies.

\section{A new paradigm for the formation of S0/dS0/dE/dSph/UCD galaxies}

Once a disky object has formed then it may enter a denser environment, whether a 
galactic, group or cluster mass system. If the velocity dispersion of the deeper
potential is more than $\sim$five times the velocity dispersion of the infalling galaxy
then it is unlikely to merge with either the central object (by dynamical friction)
or with another satellite.
Only external processes will effect its evolution and we can speculate, with support from
numerical simulations, that the entire sequence of remaining galaxies (S0, dS0, dE, dSph...)
is created by the transformation of spiral or dIrr systems by 
impulsive and resonant gravitational interactions with some additional help from 
hydrodynamical processes.

  \begin{figure}
    \centering
     \includegraphics[width=5.5cm,angle=0]{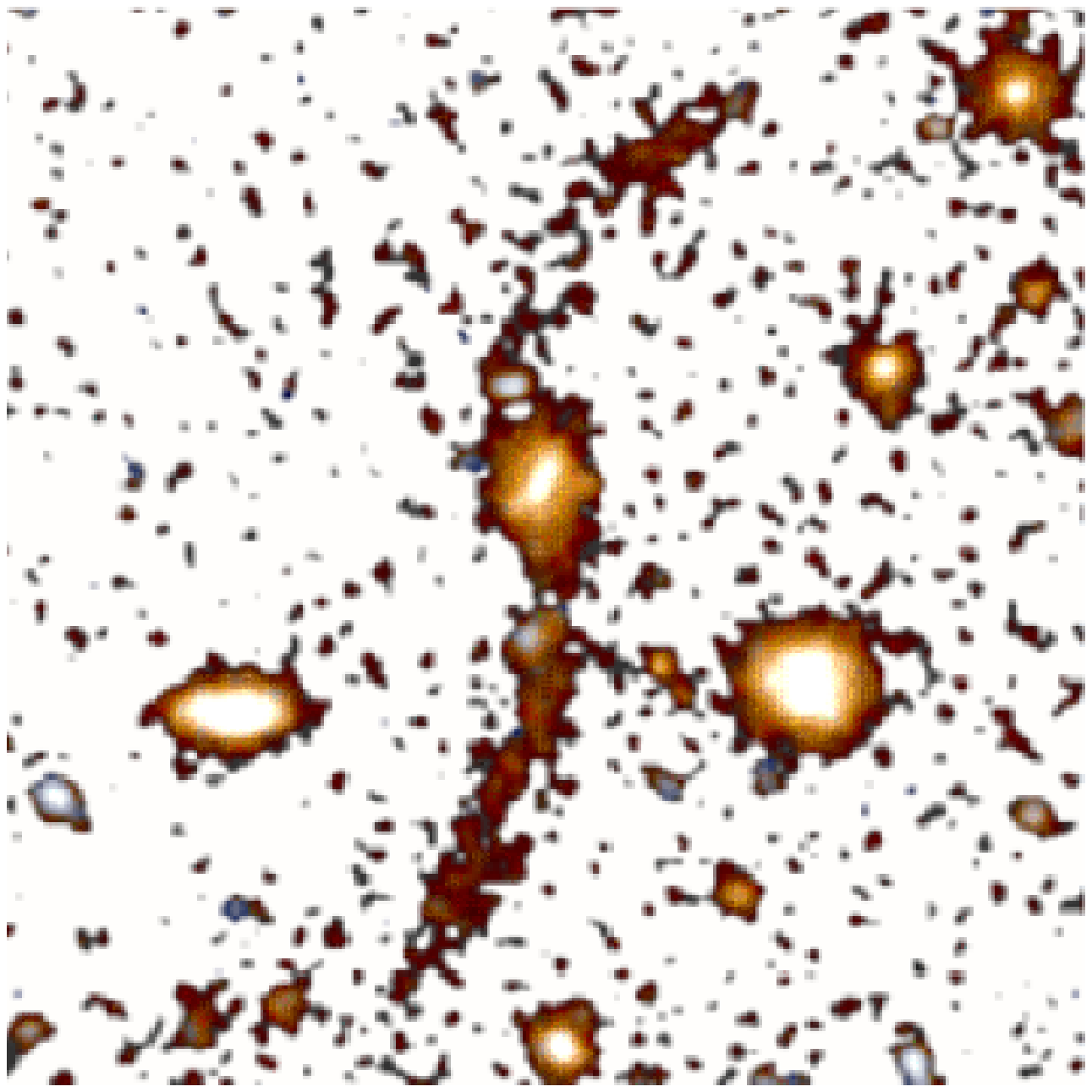}
     \includegraphics[width=5.5cm,angle=0]{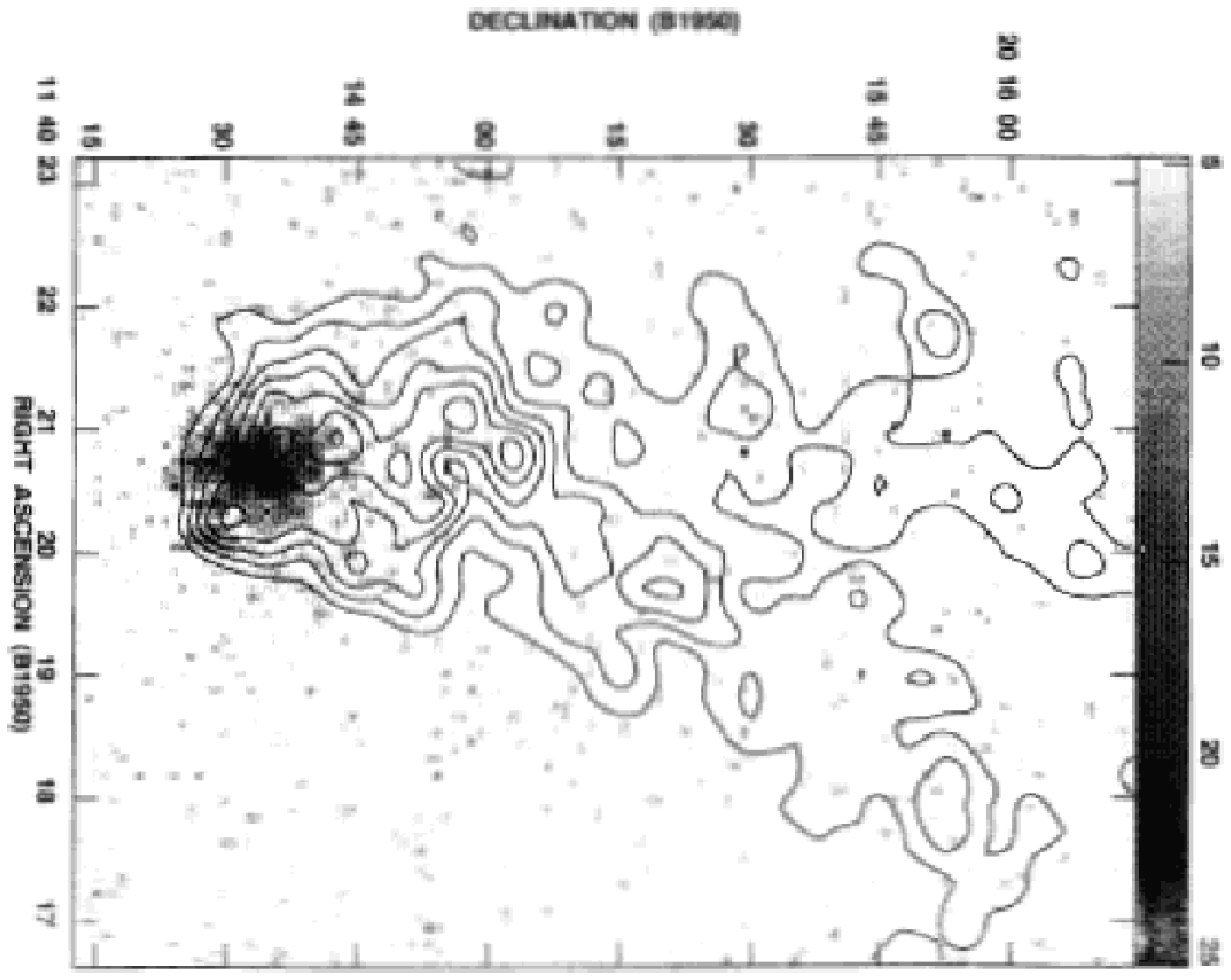}
    \caption{The left image shows a symmetric stellar tidal tail from a disk galaxy that is orbiting
in a galaxy cluster at z=0.5 (Courtesy of I. Smail and the MORPHS group). The right image shows
the trailing radio emission from the hydrodynamical stripping of a cluster galaxy 
(courtesy of G. Gavazzi). Direct observational
evidence for either gravitational or hydrodynamical effects are hard to find. In general, the tidal
debris from harassed disks is too faint to observe, and ram-pressure stripping acts on such
a short timescale that it is rare to catch a galaxy in the act of being stripped.}
    \label{sample-figure}
  \end{figure}

Ram pressure stripping is effective at removing the entire gas supply from galaxies that pass through
the cores of rich clusters (c.f. Figure 1.3). 
This will supress star-formation in cluster galaxies but will it be effective
at large distances from the centres of clusters where the gas density is low?
Gravitational interactions may be important over a larger
region of the cluster since more dark matter is bound to galaxies further from the central
cluster potential (thus the encounters are stronger) but the number
of encounters between halos decreases. 
The orbits of galaxies
in clusters are  nearly isotropic, which results in most of the galaxies orbiting through the dense
inner region of the cluster (Ghigna etal 1998). In fact 10\% of orbits will take galaxies through the
core and to beyond 
twice the cluster virial radii ($\sim$6 Mpc for Coma). Therefore the environment near rich clusters 
may host galaxies that have suffered significant gravitational perturbations and have 
been partially stripped of gas (c.f. Figure 1.2) by virtue of orbiting through the cluster core.

  \begin{figure}
    \centering
     \includegraphics[width=8cm,angle=0]{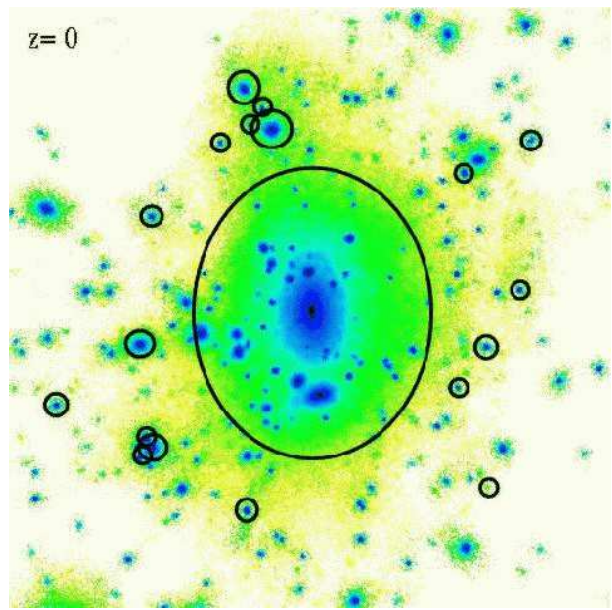}
    \caption{The density of dark matter plotted to 3 virial radii for a high resolution simulation
of a galaxy cluster. The large inner circle is the virial radius of the cluster. The smaller circles highlight 
halos that are currently outside the cluster but have orbited within $0.25r_{200}$.
(This is the final frame of the mpeg movie showing the formation history of this cluster
and can be downloaded from www.nbody.net.)
}
    \label{sample-figure}
  \end{figure} 

  \begin{figure}
    \centering
     \includegraphics[width=10cm,angle=0]{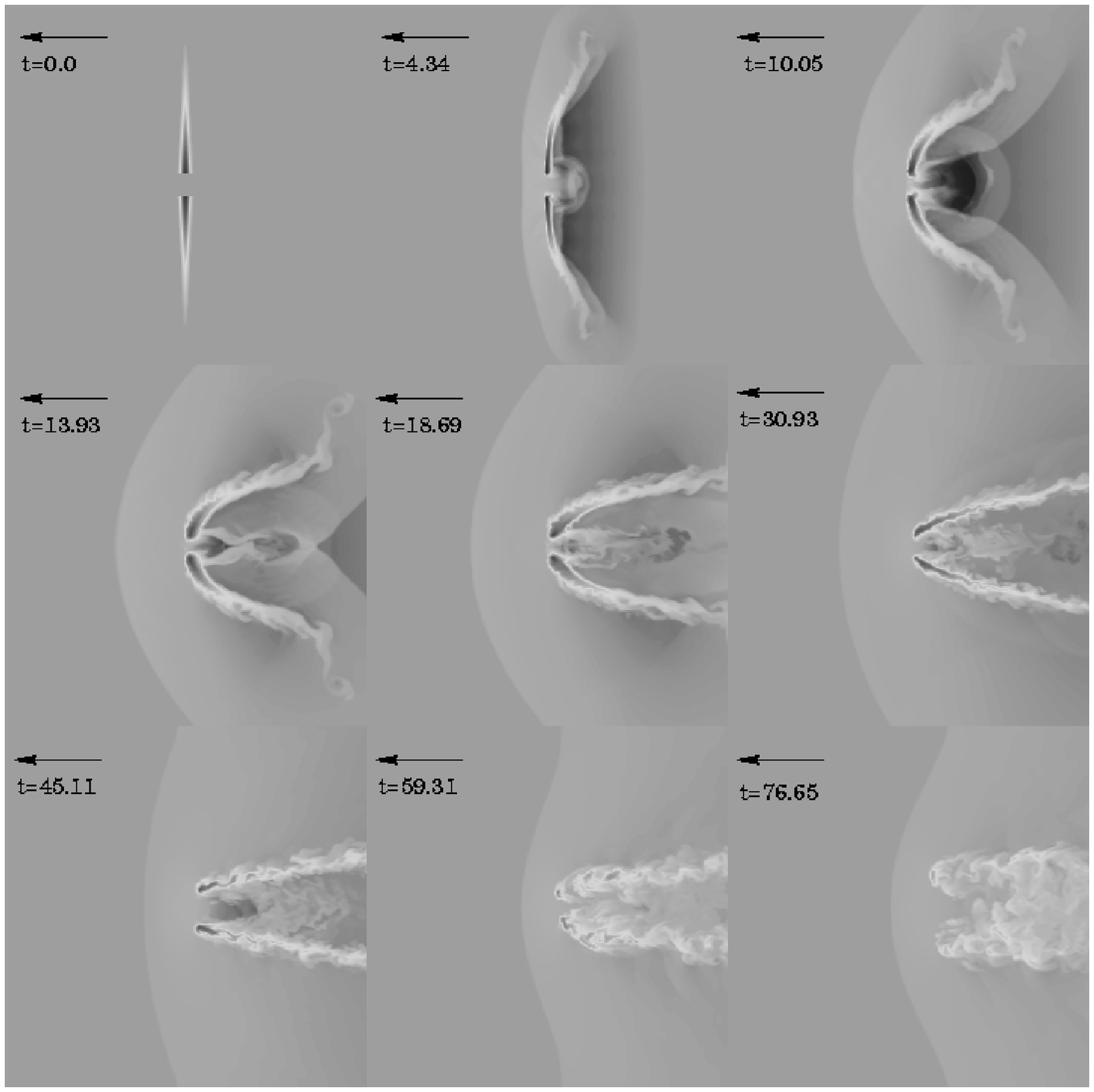}
    \caption{A high resolution Eulerian simulation of gas stripping of a galaxy falling into 
a gas density representitave of the centre of the Coma cluster at 3000 km/s (Quilis etal 2000). 
The time unit is millions of years and each frame is 50 kpc on a side.}
    \label{sample-figure}
  \end{figure}

The definition of an S0 is basically a featureless disky galaxy with little or no ISM, so the
simplest way to create an S0 is to remove the ISM from a Sa/Sb disk (Solanes etal 1989, 1992).
We have two possible methods for accomplishing this. Any galaxy passing through the core of a
massive cluster will have its gas supply completely stripped by ram-pressure and viscous stripping
on a timescale smaller than the core crossing time. The 
remaining dense molecular gas may recycle into the ISM and also be removed or
 consumed in a burst of star-formation due to the pressure 
increase as the galaxy reaches the cluster center. 
Alternatively, gravitational interactions with the cluster and its substructures will
heat disks, raising the Toomre ``Q'' parameter, which naturally suppresses spiral structure and other
instabilities in the disk, further suppressing star-formation as molecular cloud growth is halted
(Moore etal 1998, Gnedin 2003a, Gnedin 2003b). 

  \begin{figure}
    \centering
     \includegraphics[width=11cm,angle=0]{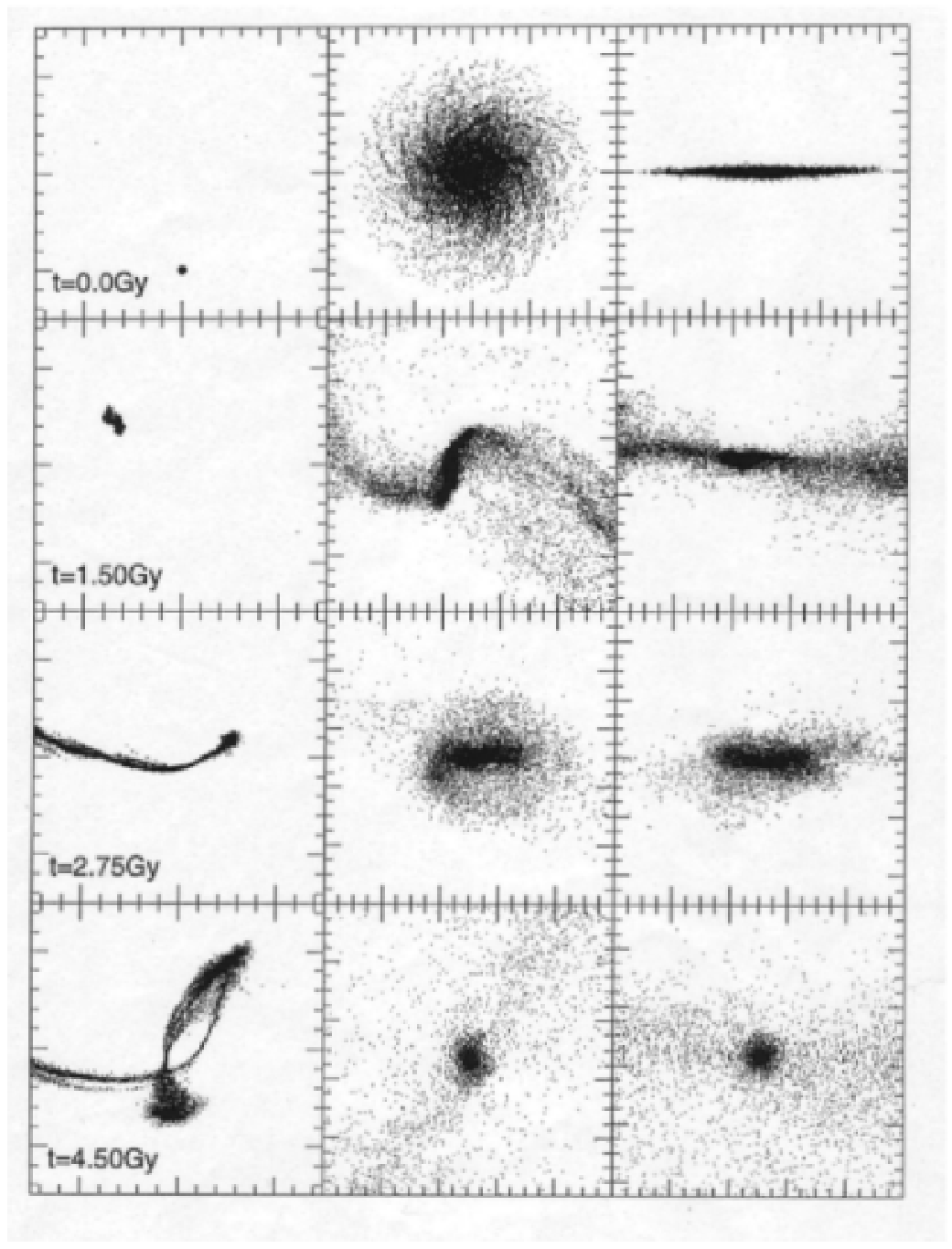}
    \caption{Gravitational tides from a deeper potential are sufficient to drive an evolution 
from disks to dwarf spheroidals. Here I show the evolution of an Sc spiral galaxy on a 6:1 (apo:peri)
orbit where the pericenter is $0.1r_{virial}$ and the ratio of circular velocities of the two systems
are 5:1. This system could therefore be rescaled to represent a galaxy like the SMC orbiting 
within the Milky Way or an $L_*$ LSB galaxy orbiting in a galaxy cluster. The left frame shows 
the entire orbit whilst the center and right frames are face and edge on views centred on the galaxy.
The forced bar instability is so violent that the evolution is better described by a secondary violent relaxation 
of the stellar disk --  most of the galaxy is stripped into symmetric debris streams leaving a small pressure 
supported spheroidal galaxy with an exponential light profile
(courtesy of C. Calcaneo-Roldan).}
    \label{sample-figure}
  \end{figure}

\subsection{Cluster dwarf ellipticals (dE's) and transition galaxies (dS0's).}

Dwarf ellipticals are the most numerous type of galaxy in clusters (Bingelli etal 1985). 
They are flattened exponential systems, sometimes with a bright compact nucleus.  
These systems are in various dynamical states ranging from pressure
supported spheroids to rotationally flattened disky systems (Geha etal 2002).
Numerical simulations of disks orbiting within cluster potentials have shown
that a dramatic transformation to a diffuse spheroidal system is 
likely to occur (Moore eta 1996).
The transformation sequence begins with a violent bar instability. Subsequent
perturbations cause the bar to 
lose angular momentum and it eventually collapses into a spheroidal system
through a buckling type instability. During the early stages of the transformation, which takes 
roughly a cluster orbital time, the morphology of the galaxy may
appear as a rotationally supported  dS0 system.

A fundamental prediction from gravitational heating mechanisms is that dE's should be
embedded within very low surface brightness tidal streams of stellar debris, as demonstrated
in Figure 1.4, and  these
streams should trace the orbital path of the progenitor galaxy. Since the relaxation
time is only short in the cluster cores these streams should survive relatively intact
at the edge of clusters, but they should become well mixed near the cluster centres.


\subsection{Intracluster diffuse light, overmerging and UCD's}

The nature of the dark matter is critical to the survival of galaxies in clusters.
If galaxies had constant density cores of just a few kpc 
(or cuspy potentials with very low concentrations) 
then they would all be easily disrupted by the cluster potential. This process is
directly analogous with the overmerging problem that was responsible for the
dissolution of subhalos in early N-body simulations (Moore etal 1996).
On the other hand, if galactic halos are cuspy and concentrated
they would all survive in clusters and we would not expect to observe a significant 
component of diffuse light. 

The Ultra Compact Dwarfs recently found in the Fornax
cluster are most likely the dense nuclei of nucleated dE galaxies that have 
been ``overmerged'' by gravitational interactions (Drinkwater etal 2000). 
These tracer cores show that the central concentrations 
of the progenitor galaxies must have been low enough such that they have
been completely disrupted by the present day. Both the fraction of diffuse light 
stripped from galaxies and
the abundance and locations of the UCD's could be used to constrain the
structure of galaxies and the efficiency of gravitational interactions in different 
environments.

\subsection{Local group dwarf spheroidals}


The Local Group dwarf spheroidals are the extreme tip of the galaxy luminosity function,
and as the shallowest potentials and faintest galaxies known they provide a strong test of
our understanding of galaxy formation (Kormendy 1989).
The Local Group morphology-density relation is similar to that of rich clusters in that 
the spheroidal galaxies with no diffuse gas are located close to the host galactic potentials
of the Milky Way and M31. Again
we can ask the question: do disks know not to form near the site of a massive potential
or were they originally disks that have been transformed to spheroidals through interactions? 
Observationally this is unclear and a difficult question to answer since we can't observe
Local Group progenitors at high redshifts. 

However, a huge amount of detailed data exists for these faint Local Group 
galaxies (e.g. Grebel 2001), which must be reproduced by a successful model for their formation.
Simulations of galaxy formation in a cosmological context that can resolve the faintest satellites
are some years away, therefore only the most basic comparisons between theoretical models and data
have been made to date.
The dynamics of these systems indicate that their spheroidal shapes are due to random
stellar motions and that rotational support is small.  Many of these systems show evidence 
for continuous star-formation, or widely separated bursts,  indicating that
ram-pressure stripping by a hot Galactic halo component has not been completely efficient at
removing their fuel. It also indicates that supernovae winds
have not been effective at ejecting the bulk of the gas from these systems.

Galaxies are on average some 5 Gyrs
older than galaxy clusters, which allows 
time for gravitational interactions to transform disks
to dSph's in the potential of a more massive system (Mayer etal 2001). A high density Draco like dSph
must have had a progenitor with a similar high dark matter density, such as
the dIrr galaxy GR8. However, one should not compare GR8 directly with Draco since
GR8 has evolved for 10 Gyr longer in the field than the system that accreted into the
Galaxy to evolve into Draco.  The morphological transformation predicted by the numerical
simulations is similar to that for cluster galaxies, but on a longer timescale 
since only the main galactic potential has been considered as the perturber. If the CDM model
is in fact correct then the transformation for the dSph's may be more rapid due to encounters
with dark matter substructure predicted to fill galactic halos.

\section{Conclusions}

Numerical simulations have shown that environmental processes are
sufficient to reproduce some of the basic properties of
a large fraction of the Hubble sequence (S0, dE,
dS0, dSph, UCD) by gravitationally interactions acting on disks. 
Hydrodynamical processes can can also play a role in suppressing 
star-formation on a shorter timescale but only in regions of high gas
density. Dramatic improvements to the modelling
will come when algorithms have improved to the extent that we can form
realistic disk systems from ab-initio initial conditions. This will
allow the morphological evolution in groups and clusters to be studied
in great detail and within a cosmological context, enabling quantitative
observational comparisons and predictions to be made.  Realistically this is probably
5--10 years away. For comparison with theory, observers need to
quantify the Butcher-Oemler effect using mass selected samples of
clusters at different epochs through a combination of lensing and high
resolution spectro-photometric data. Probing down to dwarf galaxy
luminosities and group mass scales will be of great interest to
theorists working to constrain evolutionary scenarios and cosmological
models.

\end{document}